\documentclass[aoas,nameyear,dvips]{arximspdf}
\usepackage{stfloats}
\usepackage{graphicx}

\doi{10.1214/10-AOAS358}
\volume{4}
\issue{4}
\pubyear{2010}
\firstpage{1824}
\lastpage{1846}

\makeatletter
\fnbelowfloat
\newcommand{\betab}{\bolds{\beta}}
\makeatother

\begin{document}
\begin{frontmatter}

\title{Zero-inflated truncated generalized Pareto distribution for the
analysis of radio audience data}
\runtitle{Zero-inflated truncated generalized Pareto distribution}

\begin{aug}
\author[A]{\fnms{Dominique-Laurent} \snm{Couturier}\ead[label=e1]{Dominique.Couturier@unige.ch}}
\and
\author[A]{\fnms{Maria-Pia} \snm{Victoria-Feser}\corref{}\ead[label=e2]{Maria-Pia.VictoriaFeser@unige.ch}\thanksref{t1}}

\thankstext{t1}{Supported by the Swiss National Science Foundation
Grant PP001-106465.}
\runauthor{D.-L. Couturier and M.-P. Victoria-Feser}
\affiliation{University of Geneva}
\address[A]{HEC Gen\`{e}ve\\
University of Geneva\\
40 Bd du Pont d'Arve\\
1211 Geneva 4 \\
Switzerland\\
\printead{e1}\\
\phantom{E-mail:\ }\printead*{e2}} 
\end{aug}

\received{\smonth{4} \syear{2009}}
\revised{\smonth{1} \syear{2010}}

%
\begin{abstract}
Extreme value data with a high clump-at-zero occur in many domains.
Moreover, it might happen that the observed data are either truncated
below a given threshold and/or might not be reliable enough below that
threshold because of the recording devices. These situations occur, in
particular, with radio audience data measured using personal meters
that record environmental noise every minute, that is then matched to
one of the several radio programs. There are therefore genuine zeros
for respondents not listening to the radio, but also zeros
corresponding to real listeners for whom the match between the recorded
noise and the radio program could not be achieved. Since radio
audiences are important for radio broadcasters in order, for example,
to determine advertisement price policies, possibly according to the
type of audience at different time points, it is essential to be able
to explain not only the probability of listening to a radio but also
the average time spent listening to the radio by means of the
characteristics of the listeners. In this paper we propose a
generalized linear model for zero-inflated truncated Pareto
distribution (ZITPo) that we use to fit audience radio data. Because it
is based on the generalized Pareto distribution, the ZITPo model has
nice properties such as model invariance to the choice of the threshold
and from which a natural residual measure can be derived to assess the
model fit to the data. From a general formulation of the most popular
models for zero-inflated data, we derive our model by considering
successively the truncated case, the generalized Pareto distribution
and then the inclusion of covariates to explain the nonzero proportion
of listeners and their average listening time. By means of simulations,
we study the performance of the maximum likelihood estimator (and
derived inference) and use the model to fully analyze the audience data
of a radio station in a certain area of Switzerland.
\end{abstract}

%

\end{frontmatter}

\section{Introduction}

Audience indicators---like
rating,\setcounter{footnote}{1}\footnote{Percentage of people who tune
in to a given radio station during a day.} time spent
listening\footnote{Average listening time to a given radio station per
listener.} and market share---are essential for radio stations
managers and advertisers. They give important indications on public
profiles and on radio stations benchmarking, allowing proper radio
programming and optimization of advertising strategies. The weaknesses
of traditional audience measurements methods based on individual
recollection of the time spent listening to all radio stations led to
the development of individual, portable and passive electronic
measurement systems providing more reliable and detailed measures
[refer to \citet{Websterratings2006} for a complete overview of
audience measurement methods].
Telecontrol\footnote{\url{http://www.telecontrol.ch}.} thus developed
a ``wristwatch meter,'' which records 4 seconds of ambient sound at fix time delays and compares these sequences to the corresponding ones of all
available radios. The ``people portable meter'' of
Arbitron\footnote{\url{http://www.arbitron.com}.} or the ``Eurisko
multimedia monitor'' of Gfk\footnote{\url{http://www.gfk.com}.} consist
in a pager-sized device which detects inaudible codes that broadcasters
embed in their programs.

Hence, the fundamental audience measure
available through these portable and passive measurement systems is a
dichotomous variable $Y_{ismt}$ indicating if the participant $i$ was
listening to the radio station $s$ at the measurement $m$ of the day $t$.
Most used audience indicators for a given radio station are all
functions of the sum of those quantities over a day part, mostly 24
hours, that is, $Y_{ist}=\sum_{\forall m}Y_{ismt}$.

%
\begin{figure}[b]

\includegraphics{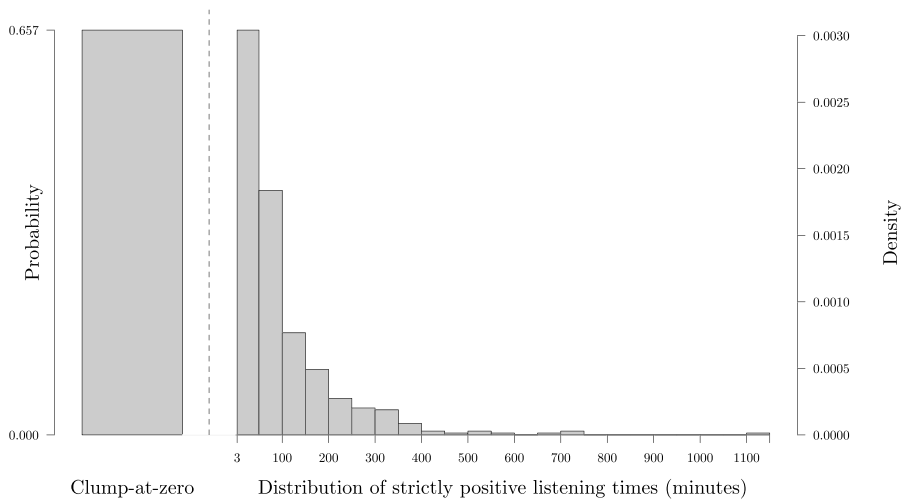}

\caption{Empirical distribution of the daily
listening times
to a national radio in an area of the French part of Switzerland during
the first semester of 2006. 1382 participants were measured by means of
the Radiocontrol system during one day of the period of interest. Zeros
represent 65.7\% of the data. The distribution of the positive data is
extremely skewed with a maximum daily listening time of 1136 minutes.
The lowest possible positive listening time is 3 minutes.}\label{fig:edf}
\end{figure}

We have at our disposal radio audience data of the Swiss measurement
system ``Radiocontrol'' in 2007 [refer to \citet{Daehler2006} for a
complete presentation of this measurement system in Switzerland]. As
illustrated in Figure~\ref{fig:edf}, the distribution of the daily
number of listening minutes $y_{ist}$ for a given radio is extremely
skewed, left-truncated and clumped-at-zero. In other words, first, the
empirical distribution of the data appears monotonically decreasing.
The probability to listen to a radio during a time interval decreases
with the time interval length. Second, because of contact validation
rules of the Swiss measurement system, the listening times $y_{ist}$
are recorded as zeros if none of the contacts of the participant $i$ to
the radio station~$s$ last 3--4 minutes or more on day $t$. This
means that the smallest observed (recorded as such) listening times are
3--4 minutes. This ensures that the probability to observe false
positive contact is negligible over a time interval of 4 or more
consecutive minutes. Third, the data contain a high clump-at-zero
corresponding to the percentage of people that had no recorded contact with that radio
station.

Data with a clump-at-zero and an asymmetric heavy-tail distribution
occur in numerous disciplines. Examples are the daily levels of
precipitation in an area [\citet{Weglarczyk05}], the yearly amount of
car insurance claims per client [\citet{Chapados02};
\citet{Christmann04}] or the length of overnight stays at hospital per
patient [\citet{Chen07}]. However, no model has been proposed so far
for data with a clump-at-zero together with a truncation of small
values under a threshold, a model that is necessary to describe, in
particular, radio audience data like in our example, but also any other
type of data that might, for example for recording reasons, have
unreliable measurements at small values of the variable of interest.
Hence, the purpose of this paper is to develop a model able to fit
truncated heavy-tailed data with excess zeros and to explain, by means
of covariates, both the probability associated with a nonzero value and
the expectation of positive outcomes. Such a model particularly makes
sense in the context of radio audience: The probability of a nonnull
value and the expectation of positive outcomes, respectively,
correspond to the rating and time spent listening audience indicators.
Market shares are a function of these expectations.

Models for data with excess zeros have received much attention in the
literature. The most popular ones include the two-part model of
\citet{DuanAll83} and the zero-inflated count models initiated by
\citet{Lambert92} for continuous data, or the hurdle model of
\citet{Mullahy86} for count data. In Section~\ref{s:model} we describe
our model as a natural extension of these models that take into account
the left truncation of the outcome variable. To model the positive part
of the radio listening times, we propose a zeromodal Pareto-like
distribution. Choice has been made for the generalized Pareto
distribution because of its ability to fit heavy tails, to be ``model
invariant'' to the choice of the threshold for the left truncation, and
because it can be used to only model the tail of the distribution. The
resulting model we propose is hence a zero-inflated truncated Pareto
(ZITPo) model in which the probability of nonzero outcomes and the mean
of the positive outcomes are linked to a set of covariates in a
generalized linear model framework. The ZITPo has great fitting
flexibility and useful properties as argued in
Section~\ref{Sec_ZITPo-extensions}. In Section~\ref{s:estimation} we
investigate by means of simulations the sample properties of the
maximum likelihood estimator and inferential procedures. Since ZITPo
models are new, it is also important to be able to check the fit of the
model and, therefore, we propose in Section~\ref{s:modelcheck} a new
data analysis tool based on Pareto residuals that is derived in a
natural manner from the properties of the ZITPo model. The data from a
radio station in a certain area of Switzerland are then fully analyzed
in Section~\ref{s:application} by means of the ZITPo which provides an
excellent fit to the data and hence good explanatory power for the
probability of nonzero outcomes and the mean of the positive outcomes.

\section{The ZITPo model\label{s:model}}

The generalized Pareto distribution, introduced by \citet{Pickand75},
is a limit distribution for the excess over a (large) threshold
$\alpha$ for data coming from generalized extreme value distributions,
as well as a generalization of the Pareto distribution. The three
parameter generalized Pareto distribution has the following
cumulative distribution function: 
%
\begin{equation}\label{eq:gpd}
F_{Y}(y|\alpha,\tau,\xi)=
\cases{\displaystyle
1-\biggl(1+\xi\frac{y-\alpha}{\tau}\biggr)^{-1/\xi}, &\quad \mbox{if } $\xi
\neq0$, \cr
\displaystyle 1-\exp\biggl(-\frac{y-\alpha}{\tau}\biggr),&\quad \mbox{if } $\xi=
0$,}
\end{equation}
where $\alpha$, $\tau$ and $\xi$ are location, scale and shape
parameters, $\alpha\geq0$ and $\tau>0$. The range of $y$ is $]
\alpha,-\frac{\tau}{\xi+\alpha}[$ if $\xi<0$, and $] \alpha
,\infty[$
otherwise. The exponential distribution with mean $\tau$ occurs for
$\xi=0$. Pareto-like distributions occur for $\xi>0$. The generalized
Pareto distribution has been widely used to model rare events in
several fields. Applications for environmental extremes are especially
numerous (river flow, ozone levels, earthquakes).

For modeling audience radio data, it is also important to be able to
link moments or parameters of the generalized Pareto distribution to a
set of explanatory variables. The generalized linear models (GLM)
framework, introduced by \citet{NelderWedd72}, provides a general
setting to achieve this aim. GLM are a generalization of the linear
regression model in which the assumption of normality of the
conditional distribution of the response vector $\mathbf{y}$ given a
set of covariates $\mathbf{X}$, $\mathbf{y}|\mathbf{X}$, is relaxed.
These models assume that the $i$th unit response, $y_{i}$, follows a
distribution belonging to the exponential family, and the expectation
of the $i$th response, $y_{i}$, is linked to a set of fixed covariates
$\mathbf{x}_{i}$ through an invertible linear predictor function
$\nu(\cdot)$, by means of
$\mathrm{E}[{Y}_{i}]=\nu^{-1}(\mathbf{x}_{i}\betab)$, with $\betab$ a set
of regression coefficients. The generalized Pareto distribution falls
outside the exponential family framework and, hence, the advantages
associated with this framework---like well-known iterative estimation
procedures and mathematical properties---are not available. However,
extension of the GLM to distributions outside the exponential family is
pretty straightforward.

Actually, generalized linear modeling has existed for a long time with
responses following extreme value distributions, but not in the
traditional scheme that directly relates the response expectation to
the explanatory variables through a linear predictor. Indeed, in
extreme value analyses, very often the parameters of the response
distribution instead of the response expectation are linked to the
covariates. Davison and Smith [(\citeyear{DavisonSmith-90}), page 395] consider that this
represents ``a more fruitful approach'' than the usual one that links
the distribution moments to the regressors, as the moments of
generalized extreme value distribution do not exist for all values of
their parameters. We refer to Coles [(\citeyear{Coles01}), Section 6.4] for a
review. In survival analysis, depending on the choice of the hazard
function $h(t)$, the survival function $f(t)$ may follow an extreme
value distribution. In this context, the hazard function
$h(t)=\frac{f(t)}{1-F(t)}$ is then related to the covariates through a
linear predictor instead of the response expectation. Such developments
may be found in \citet{AitkClay80}. As we will see in more details
below, for the purpose of modeling radio audience data, it is more
sensible to link the expected value of the response to a set of
covariates.

Before adapting the generalized Pareto distribution to handle
clump-at-zero and left truncation of the positive part of the data, as
well as incorporating in the resulting model
covariates in order to explain the probability of a zero outcome and
the mean of the positive part, we briefly describe models proposed so
far for data with excess zeros. The aim is to propose a general
formulation from which different models for different situations can be
deduced, and, in particular, from which we build our zero-inflated
truncated Pareto (ZITPo) model. We then also describe in details the
ZITPo model assumptions and discuss some possible extensions.

%
%

\subsection{Models for nonnegative data with excess zeros}\label{Sec_ZI-models}

There is a rich literature about adaptation of statistical models to
the case of data with excess zeros. We refer to Min and Agresti
(\citeyear{MinAgresti02}, \citeyear{MinAgrerand2005}) and \citet{ridout98} for a review.
\citet{MinAgresti02} compare the advantages and disadvantages of
existing approaches and note that the most appealing modeling for
continuous data with excess zeros is the two-part model of
\citet{DuanAll83}, and the zero-inflated count models initiated by
\citet{Lambert92} or the hurdle model of \citet{Mullahy86} in the case
of count data with a clump-at-zero. These models are similar. Their key
idea is to mix two random variables: A first one, $Y_{1}$, that handles
the excess of zeros, and a second one, $Y_{2}$, that models the other
part of the data. $Y_{1}$ typically follows a Bernoulli distribution
where $P_{Y_{1}}(0)=1-\pi$ denotes the probability to observe a zero
outcome. In the hurdle and two-part models (also called conditional
models), the probability of the data being equal to zero only depends
on $Y_{1}$ and the positive data are all modeled by $Y_{2}$, which may
follow a zero-truncated distribution in the case of count data (hurdle
model) or a continuous distribution (two-part model). In these cases,
$P_{Y_{2}}(0)=0$. In zero-inflated models (also called mixture models),
$Y_{2}$ does not follow a zero-truncated distribution. The probability
associated to zero thus depends on both $Y_{1}$ and $Y_{2}$.

Let $Y$ be a random variable with probability distribution $P_{Y}$ for
the clump-at-zero and the positive part, when the latter is discrete,
that is, $Y_{2}$ is discrete, then $P_{Y}$ may be expressed in the
following way:
\begin{eqnarray}\label{eq:zerodiscr}
P_{Y}(y) &=&\bigl[P_{Y_{1}}(0)+\bigl(1-P_{Y_{1}}(0)\bigr) P_{Y_{2}}(y)\bigr]\iota(y=0)
\nonumber\\[-8pt]\\[-8pt]
&&{}+\bigl[\bigl(1-P_{Y_{1}}(0)\bigr) P_{Y_{2}}(y)\bigr]\iota(y>0)\nonumber ,
\end{eqnarray}
where $y=0, 1, 2,\ldots,$ the indicator function $\iota(\cdot)$ equals
one if the condition is true and zero otherwise. Let us refer to a
variable as semicontinuous when it has a point mass in zero and a
continuous distribution for the positive values [definition of \citet{MinAgresti02}, page 7]. Then (\ref{eq:zerodiscr}) may easily be
generalized to continuous or semicontinuous $Y_{2}$:
\begin{eqnarray}\label{eq:zerocont}
f_{Y}(y) & =& \bigl[P_{Y_{1}}(0)+\bigl(1-P_{Y_{1}}(0)\bigr) P_{Y_{2}}(0)\bigr]\delta(y)\nonumber\\[-8pt]\\[-8pt]
&&{}+ \bigl[\bigl(1-P_{Y_{1}}(0)\bigr) f_{Y_{2}}(y) \bigr]\Delta_{0}(y)\nonumber,
\end{eqnarray}
where $\delta(y)$ is a Dirac delta function which equals zero
for $y \neq0$, $\Delta_{0}(y)$ is a step function taking the value of
one for $y>0$ and zero otherwise, and $y\in[0,\infty[$. Note that when
$P_{Y_{2}}(0)=0$, we have the hurdle or two parts models, while we have
zero-inflated models when this is not the case.

The use of the generalized Pareto distribution to model zero-inflated
data is not common, one exception being \citet{Weglarczyk05}. The
authors compare the fitting ability of some semicontinuous
distributions to fit hydrological data with excess zeros and consider a
Dirac generalized Pareto distribution with density function
%
\begin{equation}\label{eq:singh}
f_{Y}(y|\pi,\tau,\xi)=(1-\pi)\delta(y)+\frac{\pi}{\tau}\biggl(1+\xi
\frac{y}{\tau}\biggr)^{-1/\xi-1}\Delta_{0}(y),
\end{equation}
where $\tau>0$, $\xi\neq0$, $0\leq(1-\pi)\leq1$ corresponds to the
probability of a zero event. Note that compared to (\ref{eq:gpd}),
$\alpha=0$. The Dirac generalized Pareto distribution in
(\ref{eq:singh}) thus corresponds to a two-part model with
$P_{Y_{2}}(0)=0$, in which $f_{Y_{2}}(y)$ is the density function of
the generalized Pareto distribution.

In the following sections we propose to extend (\ref{eq:zerocont})
[and (\ref{eq:singh})] to take into account the possible truncation of
small values, as well as to incorporate
covariates to explain (a function of) the probability of zero outcomes
and the mean distribution of positive outcomes.

\subsection{The ZITPo distribution}\label{Sec_ZITPo}

Let $Y^{*}$ denote the effective (but unknown) daily listening time for
a given radio. $Y^{*}$ is to the sum over the day of the dichotomous
variables indicating a contact to that radio station minute by minute.
The probability and cumulative distribution functions of $Y^{*}$,
$f_{Y^{*}}(y^{*})$ and $F_{Y^{*}}(y^{*})$, are semicontinuous with a
point mass in zero and a continuous distribution for the positive
values. Let $Y$ denote the observed listening times with density
function $f_{Y}(y)$. As listening times smaller than a given value
$y^{\circ}$ (considered as known) are recorded as zeros, observed zeros
are then a mixture between the effective zero listening times and the
positive listening times reported as zeros because of the measurement
system. Accordingly, $F_{Y}(0)=F_{Y^{*}}(y^{\circ})$.

A semicontinuous version of the zero-inflated count model described in
(\ref{eq:zerocont}) is indeed adequate to model the double origins of
the zeros in the clump-at-zero and the positive values of the observed
listening times. Let us assume that the unknown and true proportion of
zero listening times is $1-\pi$, with $0\leq\pi\leq1$, and that the
effective positive listening times follow a two parameter generalized
Pareto distribution (with $\alpha=0$), $Y^{*}|(Y^{*}>0)\sim\mathrm{GPD}(\tau,\xi)$. Then, in (\ref{eq:zerocont}), $P_{Y_{1}}(0)=1-\pi $
corresponds to the effective proportion of nonlisteners, and
$P_{Y_{2}}(0)=F_{(Y^{*}|Y^{*}>0)}(y^{\circ})$ corresponds to the part
of the two parameter generalized Pareto distribution that cannot be
observed because of the measurement system limitations. The density
functions of the effective listening times $Y^{*}$ and of the observed
listening times $Y$ are

%
\begin{eqnarray}\label{eq:semigpdtau2}
f_{Y^{*}}(y^{*}|\pi,\tau,\xi) &= &[ 1-\pi]\delta(y^{*}) + \biggl[
\frac{\pi}{\tau} \biggl(1+\xi\frac{y^{*}}{\tau} \biggr)^{-1/\xi-1}
\biggr]\Delta_{0}(y^{*}),\\
f_{Y}(y|\pi,\tau,\xi) & =& \bigl[ (1-\pi) + \pi
F_{(Y^{*}|Y^{*}>0)}(y^{\circ}) \bigr]\delta(y)\nonumber\label{eq:semigpdtau}\\
&&{}+ \bigl[ \pi f_{(Y^{*}|Y^{*}>0)}(y)\bigr]\Delta_{y^{\circ}}(y)
\nonumber\\[-8pt]\\[-8pt]
& = &\biggl[ 1-\pi\biggl(1+\xi\frac{y^{\circ}}{\tau}\biggr)^{-1/\xi} \biggr]\delta(y)\nonumber\\
&&{} +
\biggl[ \frac{\pi}{\tau} \biggl(1+\xi\frac{y}{\tau} \biggr)^{-1/\xi-1} \biggr]\Delta_{y^{\circ}}(y)\nonumber,
\end{eqnarray}
where $0\leq\pi\leq1$, $\tau>0$, $\xi\neq0$ and $y^{\circ}\geq0$. For
$y^{\circ}=0$, (\ref{eq:semigpdtau}) reduces to the Dirac generalized
Pareto described in (\ref{eq:singh}). Finally, note that if the
observed listening times distribution in (\ref{eq:semigpdtau}) has the
disadvantage of being a mixture distribution which makes it more
complex to fit, its underlying distribution in (\ref{eq:semigpdtau2})
takes the advantages of the orthogonal parameterization of the hurdle
and two-part models and is thus easier to interpret [for a discussion
on the orthogonal parameterization see, e.g.,
\citet{WelshCunnDonnLind1996}]. Indeed, the zeros depend on $\pi$,
while the positive outcomes rely on the generalized Pareto parameters,
$\tau$ and $\xi$.

\begin{figure}

\includegraphics{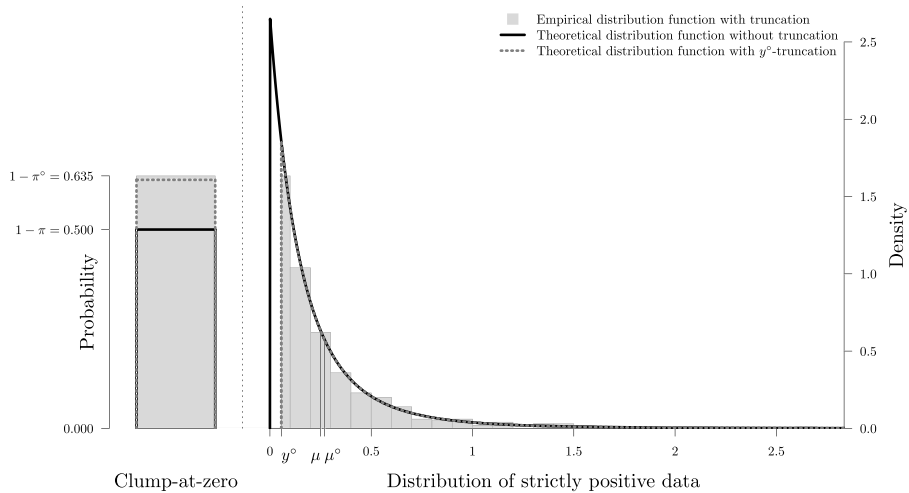}

\caption{Empirical distribution function of a data set
simulated from a ZITPo model with parameters $\pi=0.5$, $\mu=\xi=0.25$
and $y^{\circ}=F^{-1}_{(Y^{*}|Y^{*}>0)}(0.25)$. The theoretical
truncated and untruncated density functions are superimposed to the
plot with dashed gray and black lines. The value of the expectations of
the positive values of the truncated and untruncated distributions,
$\mu^{\circ}$ and $\mu$, are indicated on the x-axis. On the discrete
part of the plot, the surfaces within the dashed gray and black boxes
correspond to the theoretical probabilities to observe zeros when there
is (dashed gray) and when there is no (black) left truncation of the
positive part of the data. Those probabilities respectively equal
$1-\pi$ and $1-\pi^{\circ}=(1-\pi)+\pi
F^{-1}_{(Y^{*}|Y^{*}>0)}(y^{\circ})$.}\label{fig:semigpd}
\end{figure}

Figure~\ref{fig:semigpd} shows the distribution of a data set
simulated from a ZITPo distribution. The theoretical untruncated and
truncated distribution functions, respectively corresponding to (\ref
{eq:semigpdtau2}) and (\ref{eq:semigpdtau}), are respectively
superimposed to the plot in black and dashed gray lines. On the
discrete part of the plot, the surfaces within the dashed gray and
black boxes correspond to the theoretical probabilities to observe
zeros when there is (dashed gray) and when there is no (black) left
truncation of the positive part of the data. Those probabilities
respectively equal $1-\pi$ and $1-\pi^{\circ}=(1-\pi)+\pi
F^{-1}_{(Y^{*}|Y^{*}>0)}(y^{\circ})$. On the continuous part of the
plot, the expectations of the truncated ($\mu^{\circ}$) and
untruncated ($\mu$) distributions are indicated. It is then clear that
the expected value for the true listening time $Y^{*}$, $\mu$, is
different from the expected value of the truncated distribution, $\mu
^{\circ}$. For the audience data, one quantity of interest is $\mu$
for the untruncated distribution.

\subsection{Covariates modeling in ZITPo distribution}

Adaptation of the GLM to models for data with excess zeros is very
intuitive. The expectations of the distributions of $Y_{1}$ and $Y_{2}$
in (\ref{eq:zerodiscr}) and (\ref{eq:zerocont}) are linked to the
covariates through adapted link functions. The logit link is often
chosen to relate the expectation of $Y_{1}$, corresponding to the
probability to observe positive values, to the covariates. The log link
makes sense to connect the expectation of $Y_{2}$, corresponding to the
mean of the positive data, to the covariates, as this last is
necessarily positive. For the $i$th observation, we then have
%
\begin{eqnarray}
\pi_{i}&= & \mathrm{P}(Y^{*}_{i}>0) =
\nu_{1}^{-1}(\mathbf{x}_{i1}^{T}\betab_{1}) =
\frac{\exp(\mathbf{x}_{i1}^{T}\betab_{1})}{1+\exp(\mathbf{x}_{i1}^{T}\betab_{1})},
\label{eq:exppi} \\
\mu_{i}&= & \mathrm{E}[Y^{*}_{i}|Y^{*}_{i}>0] =
\nu_{2}^{-1}(\mathbf{x}_{i2}^{T}\betab_{2}) =
\exp(\mathbf{x}_{i2}^{T}\betab_{2}), \label{eq:expmu}
\end{eqnarray}
where $\nu_{1}^{-1}(\cdot)$ and $\nu_{2}^{-1}(\cdot)$ are the inverse
of the linear predictor functions linking the expectations of $Y_{1}$
and $Y_{2}$ in (\ref{eq:zerodiscr}) and (\ref{eq:zerocont}) to the
covariates, $\mathbf{x}_{i1}$ and $\mathbf{x}_{i2}$ are the covariates
of the $i$th observation that may contain the same predictors, and
$\betab_{1}$ and $\betab_{2}$ are the corresponding parameters. Because
of the orthogonal parameterization of the underlying model in
(\ref{eq:semigpdtau2}), if we use in (\ref{eq:exppi}) and
(\ref{eq:expmu}) two different and uncorrelated sets of covariates,
$\mathbf{X}_{1}$ and $\mathbf{X}_{2}$, we then assume that the
processes that explain the probability to observe a positive outcome
and the expectation of a positive outcome are independent. If part of
the covariates of $\mathbf{X}_{1}$ are present in (or correlated to)
$\mathbf{X}_{2}$, $\pi_{i}$ and $\mu_{i}$ will possibly be linked. No
assumption is done about the form of the relationship between these
quantities.

Inclusion of covariates in (\ref{eq:singh}) requires that we express the
distribution $f_{Y}(y)$ in terms of the expectation of the positive
values of the data. Let $(Y^{*}|Y^{*}>0)\sim\mathrm{GPD}(\tau,\xi)$.
Then
\[
\mu=\mathrm{E} [Y^{*}|Y^{*}>0 ]=\frac{\tau}{1-\xi}\qquad\mbox{for }1-\xi>0.
\]
The first moment of the generalized Pareto distribution,
$\mu$, thus exists for values of $\xi$ lower than one. Substituting
$\tau$ by $\mu(1-\xi)$ in (\ref{eq:semigpdtau}) gives
\begin{eqnarray}\label{eq:semigpdmu}
f_{Y}(y|\pi,\mu,\xi)&=& \biggl[ 1-\pi\biggl(1+\biggl(\frac{\xi}{1-\xi}
\biggr)\frac{y^{\circ}}{\mu
}\biggr)^{-1/\xi}\biggr]\delta(y) \nonumber\\[-8pt]\\[-8pt]
&&{} + \biggl[ \frac{\pi}{\mu(1-\xi)} \biggl(1+ \biggl(\frac{\xi}{1-\xi} \biggr)\frac{y}{\mu}
\biggr)^{-1/\xi-1} \biggr]\Delta_{y^{\circ}}(y) \nonumber,
\end{eqnarray}
with $0\leq\pi\leq1$, $\mu>0$, $\xi\neq0$ and $\xi< 1$,
$y^{\circ}\geq0$. The inclusion of the covariates as described in
(\ref{eq:exppi}) and (\ref{eq:expmu}) is now straightforward. For the
$i$th observation, we have
\begin{eqnarray}\label{eq:semigpdmui}
&& f_{Y_{i}}(y_{i}| \mathbf{x}_{i1},\mathbf{x}_{i2},\betab_{1},\betab_{2},\xi)\nonumber\\
&&\qquad =\,\biggr[ 1-\frac{\exp(\mathbf{x}_{i1}^{T}\betab_{1})}{1+\exp(\mathbf{x}_{i1}^{T}\betab_{1})}\biggl(1+\biggl(\frac{\xi}{1-\xi}\biggr) \frac{y^{\circ}}{ \exp(\mathbf{x}_{i2}^{T}\betab_{2})}\biggr)^{-1/\xi} \biggr]\delta(y)\nonumber
\\[-8pt]\\[-8pt]
&&\qquad\quad{}+ \biggl[
\frac{\exp(\mathbf{x}_{i1}^{T}\betab_{1})}{1+\exp(\mathbf{x}_{i1}^{T}\betab_{1})}
\frac{1}{\exp(\mathbf{x}_{i2}^{T}\betab_{2})(1-\xi)} \nonumber\\
&&\qquad\qquad\hspace*{5pt}{}\times \biggl(1+\biggl(\frac{\xi}{1-\xi}\biggr) \frac{y_{i}}{\exp(\mathbf{x}_{i2}^{T}\betab_{2})}\biggr)^{-1/\xi-1} \biggr] \Delta_{y^{\circ}}(y) \nonumber.
\end{eqnarray}

%
%

\subsection{Assumptions of ZITPo models\label{Sec_ZITPo-assumptions}}
The form of the ZITPo model implies a number of assumptions on the
distribution of the positive values:

First, the unobserved positive listening times belonging to the range
$]0,y^{\circ}[$
correspond to the nonobserved part of a left-truncated generalized
Pareto distribution. As the generalized Pareto density function is zero
modal and monotonically decreasing, this assumption implies that,
conditionally on the covariates, the probability of positive listening
times in the interval $]0,y^{\circ}[$ is higher than in any other
interval of the same size. As zapping through radio is frequent, we
believe that this assumption is realistic.

\begin{figure}[b]

\includegraphics{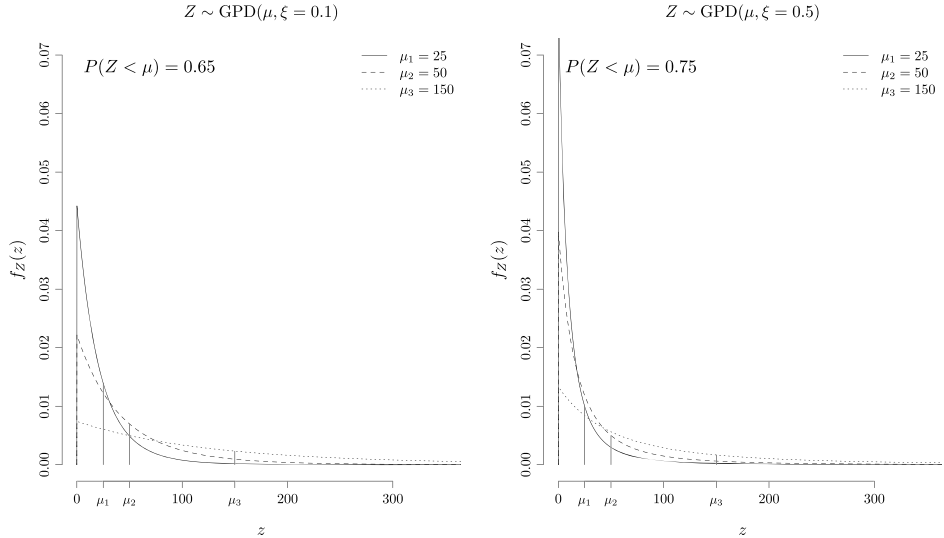}

\caption{Examples of two parameter generalized Pareto
distributions. In both plots, three distribution functions sharing the
same $\xi$-value are proposed. Their respective expectations are
$\mu_{1}=25, \mu_{2}=50$ and $\mu_{3}=100$. The probability to observe
data below the expectation is indicated above.}\label{fig:expgpd}
\end{figure}

Second, the expectation $\mu_{i}$ always corresponds to the quantile
$1-(1-\xi)^{1/\xi}$ of a $\mathrm{GPD}(\mu_{i},\xi)$. Indeed,
conditionally on the covariates, as the real positive listening times
follow generalized Pareto distributions having different expectations
$\mu_{i}$ but sharing the same $\xi$-value, $Y_{i}^{*}|(Y_{i}^{*}>0)
\sim\mathrm{GPD}(\mu_{i},\xi)$, we can observe that
%
\begin{equation}\label{eq:expgpd}
F_{(Y_{i}^{*}|Y_{i}^{*}>0)}(\mu_{i})=1-\biggl(1+\xi\frac{\mu_{i}}{\mu
_{i}(1-\xi)}\biggr)^{-1/\xi}=1-(1-\xi)^{1/\xi}.
\end{equation}
Figure~\ref{fig:expgpd} shows examples of two parameter generalized
Pareto density functions sharing the same $\xi$-value (within the same
graph) but having different expectations. For the same $\xi$-value, the
density functions show a great variety of forms and thus a high ability
to model different data sets with more or less heavy tails.

Third, because of the reparametrization of the generalized Pareto
density formulated in (\ref{eq:semigpdmu}), the shape parameter is
restricted to values lower than one. This does not seem problematic in
regard to (\ref{eq:expgpd}). Indeed, for $\xi>0.95$, $\mu$
corresponds to quantiles of the distribution higher than 0.95. We do
not
expect cases in which the theoretical mean belongs to the last 5\% of
the distribution at least with radio listening data.

Fourth, because of the logit link used in (\ref{eq:exppi}), the
probability to tune into a given radio station conditional on
covariates never equals zero or one as $\exp(\mathbf{x}_{i1}^{T}\betab_{1})>0$. We do believe that it is reasonable to
state that $0<\pi_{i}<1$ in radio audience data:
\begin{itemize}
\item As radio stations broadcast almost everywhere (airports,
supermarkets, petrol stations$,\ldots$), it seems reasonable to state that
the probability of contact of anybody is greater than zero.
\item As
radio stations do not broadcast everywhere, it also seems reasonable to
state that even the biggest fan of a specific radio station can, for
example, be outside the broadcasting range at some specific times.
\end{itemize}

\subsection{Properties and extensions of ZITPo models}\label{Sec_ZITPo-extensions}
Even if there are some restrictions in the
use of ZITPo models, the two-part form of the density described in
(\ref{eq:semigpdmui}) as well as properties of the generalized Pareto
distribution offer to ZITPo models additional abilities to fit and
analyze a variety of data sets, in particular, our radio audience data
in Switzerland:

First, one interesting property of ZITPo models is that $y^{\circ}$
may be chosen such that the observed data lower than $y^{\circ}$
integrate the most part of the false zero and false positive
observations if the data are not completely reliable in the
neighborhood of the truncation boundary. If all observed positive data
inferior to $y^{\circ}$ are coded as zeros in order to belong to the
clump-at-zero in (\ref{eq:semigpdtau}), the model will estimate the
parameters of $f_{Y^{*}}(y^{*}|\pi,\tau,\xi)$ without being affected by
the errors of the measurement system occurring on $[0,y^{\circ}[$.

Second, the stability with respect to excess over threshold operations
of the generalized Pareto distribution [see, e.g.,
\citet{CastHadi97}, page 1610, or \citet{Coles01}, page 79] and the
shifting property of distribution of the location family allow to
easily determine the distribution of the data over a threshold
$y^{\bullet}$. Let $Y_{i}^{*+}=(Y_{i}^{*}|Y_{i}^{*}>0)$ denote the
positive values of the model, with $Y_{i}^{*+} \sim\mathrm{GPD}(\tau
_{i},\xi)$ with $\tau_{i}=\mu_{i}(1-\xi)$. Then we have that
%
\begin{equation}\label{eq:distriboverbullet}
f_{(Y_{i}^{*+}|Y_{i}^{*+}>y^{\bullet})}(y_{i}^{*+}|\tau_{i},\xi) =
\frac{1}{\tau_{i}-\xi y^{\bullet}} \biggl(1+\xi
\frac{y_{i}^{*+}-y^{\bullet}}{\tau_{i}-\xi y^{\bullet}} \biggr)^{-1/\xi-1}.
\end{equation}
This is of particular interest for radio station managers and
advertisers, as important radio listeners represent the core of their
audience. The distribution of the listening time over a threshold thus
follows a three parameter generalized Pareto distribution of parameters
$\alpha^{\bullet}=y^{\bullet}$, $\tau_{i}^{\bullet}=\tau_{i}-\xi
y^{\bullet}$ and $\xi^{\bullet}=\xi$.
The corresponding expected listening time over a threshold of
$y^{\bullet}$ minutes, $\mu^{\bullet}_{i}$, is then given by
%
\begin{equation}\label{eq:muoverbullet}
\mu^{\bullet}_{i}=\mathrm{E}[Y^{*+}|Y^{*+}>y^{\bullet}]=\frac{\tau
_{i}^{\bullet}}{1-\xi}+y^{\bullet}=\mu_{i}+\frac{\xi
y^{\bullet}}{1-\xi}+y^{\bullet},
\end{equation}
where $\mu_{i}=\mu$ in simple models without covariates and
$\mu_{i}=\exp(\mathbf{x}_{i2}^{T}\betab_{2})$ in models incorporating
covariates. The expectation of the positive data over a threshold
(i.e., the expectation of the data on $]y^{\bullet},\infty[$) thus
simply corresponds to a linear shift of the expectation of the positive
data on $]0,\infty[$. There is therefore no need to change the ZITPo
model when one is interested in $\mu^{\bullet}_{i}$ or, in other words,
the effect of the covariates on $\mu^{\bullet}_{i}$ is the same as on
$\mu_{i}$.

Third, the ZITPo model can easily be extended to the three parameter
generalized Pareto distribution by introducing a shift parameter
$y^{\bullet} \leq y^{\circ}$ corresponding to $\alpha$ in (\ref
{eq:gpd}). In (\ref{eq:semigpdmu}) and (\ref{eq:semigpdmui}) we have
that $y^{\bullet}=0$. Adding the shift parameter makes sense if
information below $y^{\bullet}$ is not of direct interest, like if
nonlisteners and listeners that only zap through a given radio are
considered alike for the radio broadcaster. The resulting model which
extends (\ref{eq:semigpdmu}) [and consequently (\ref{eq:semigpdmui})]
would allow to model the probability to get an outcome lower than a
given positive value $y^{\bullet}$ as well as the expectation of the
data over $y^{\bullet}$, with positive outcomes observed above
$y^{\circ}$. In this case, all data lower than $y^{\bullet}$ would be
treated as ``zeros'' in order to be part of the clump-at-zero. The
density functions of the observed listening times $Y$ would then be
(for an observation $y_i$)
%
\begin{eqnarray}\label{eq:semigpdmui3gpd}
f_{Y}(y_i|\pi^{\bullet}_{i},\mu^{\bullet}_{i},\xi^{\bullet}) &= &\biggl[
1-\pi^{\bullet}_{i}\biggl(1+\biggl(\frac{\xi}{1-\xi}\biggr)\biggl(\frac{y^{\circ
}-y^{\bullet}}{\mu^{\bullet}_{i}-y^{\bullet}}\biggr)\biggr)^{-1/\xi}
\biggr]\delta(y_i)\nonumber \\
&&{}+ \biggl[ \frac{\pi^{\bullet}_{i} }{(\mu^{\bullet
}_{i}-y^{\bullet})(1-\xi)} \\
&&\quad\hspace*{5pt}{} \times \biggl(1+\biggl(\frac{\xi}{1-\xi}\biggr)\biggl(\frac{y_i-y^{\bullet}}{\mu^{\bullet
}_{i}-y^{\bullet}} \biggr)\biggr)^{-1/\xi-1}\biggr]\Delta_{y^{\circ}}(y_i)\nonumber.
\end{eqnarray}
The parameters $\pi^{\bullet}_{i}$ and $\mu^{\bullet}_{i}$ can possibly
be linked to a set of covariates as is done in (\ref{eq:semigpdmui}).
If there is no $y^{\circ}$-truncation and if the data on $]y^{\bullet},
y^{\circ}[$ are reliable, $y^{\circ}= y^{\bullet}$ and
(\ref{eq:semigpdmui3gpd}) is reduced to a two-part model since the
first part of the right-hand side of (\ref{eq:semigpdmui3gpd}) reduces
to $( 1-\pi^{\bullet}_{i})\delta(y_i)$. This extension is particularly
useful when the interest only lies on the tail distribution of the
positive outcomes. Indeed, in that case $\pi^{\bullet}$ is a nuisance
parameter and the generalized Pareto distributional assumption on
$]0,y^{\bullet}[$ is no more necessary. For the model to fit the data
(observed above $y^{\circ}$), one only needs the assumption that the
generalized Pareto distribution holds above $y^{\bullet}$, with a mean
that possibly depends on a set of covariates and constant $\xi$. This
might be an interesting setting, for example, in finance when seeking
to explain the value-at-risk of financial instruments. In these cases,
however, the choice of $y^{\bullet}$ might become an important issue
and criteria based on mean squared errors [see, e.g., \citet{Hill1975};
\citet{HallWelsh1985b}; \citet{BeirlantVyncTeug1996b}] or prediction
errors [\citet{DupuisVFeser2006}] could, in principle, be extended to
the ZITPo. In what follows, we will, however, focus on models with
$y^{\bullet}=0$.

\section{Estimation and inference}\label{s:estimation}

Fitting methods for the generalized Pareto distribution in
(\ref{eq:gpd}) (i.e., without a clump-at-zero) has been of great
interest in the literature. \citet{CastHadi97} and \citet{SingAhma04}
propose a comparative evaluation of the most used classical estimators
for the two and three parameter distributions. Robust estimators have
also been developed [\citet{DupuisTsao1998}; \citet{PengWelsh2001};
\citet{JuarezSchucany2004}]. We propose here to use the maximum
likelihood estimator (MLE).

The log-likelihood function of the ZITPo model described in (\ref
{eq:semigpdmui}) is
\begin{eqnarray}
&&l(\betab_{1},\betab_{2},\xi|\mathbf{y},y^{\circ},\mathbf{X}_{1},\mathbf{X}_{2})\nonumber\\
&&\qquad=\Biggl\{ \sum_{i=1}^{n} \iota(y_{i}=0) \log\biggl[
1-\frac{\exp(\mathbf{x}_{i1}^{T}\betab
_{1})}{1+\exp(\mathbf{x}_{i1}^{T}\betab_{1})}\nonumber\\
&&\hspace*{132pt}{} \times \biggl(1+ \biggl(\frac{\xi}{1-\xi}\biggr)\biggl(\frac{y^{\circ}}{
\exp(\mathbf{x}_{i2}^{T}\betab_{2})}\biggr)\biggr)^{-1/\xi}\biggr]
\Biggr\} \\
&&\qquad\quad{}+  \Biggl\{ \sum_{i=1}^{n} \Delta_{y^{\circ}}(y)
\biggl[\mathbf{x}_{i1}^{T}\betab_{1}- \mathbf{x}_{i2}^{T}\betab_{2} -
\log\biggl(\frac{(1-\xi)^{-1} }{1+\exp(1+\mathbf{x}_{i1}^{T}\betab
_{1})}\biggr) \biggr]\Biggr\} \nonumber\\
&&\qquad\quad{}+  \Biggl\{ \sum_{i=1}^{n} \Delta_{y^{\circ}}(y) \biggl(-\frac{1}{\xi}-1 \biggr)\log
\biggl(1+\biggl(\frac{\xi}{1-\xi}\biggr)\biggl(\frac{y_{i}}{\exp(\mathbf{x}_{i2}^{T}\betab_{2})}\biggr) \biggr)\Biggr\}\nonumber.
\end{eqnarray}
Maximization of this expression is achieved using the quasi-Newton
method with a numerically computed gradient matrix. Convergence is
obtained rapidly for most of the cases we have tried, even with models
embedding many covariates. The use of slightly different starting
values did always provide a solution to the unusual cases in which we
met convergence problems. The program is implemented in R functions
available in \citet{DLCMPVF10}.


In order to check the finite sample properties of the MLE for the ZITPo
model, we perform a simulation study of models incorporating covariates
as in (\ref{eq:semigpdmui}). The MLE is computed on samples with three
different sample sizes of respectively 500, 1000 and 2000 observations,
simulated with two different values for the shape parameter, $\xi
=0.25$ and $\xi=0.5$. The sampling distribution of the MLE are
presented by means of boxplots on 2500 simulated data sets. Horizontal
gray lines indicate the position of the true parameter values. The
coverage levels of 95\%-confidence
intervals of the form $[\hat{\theta}-\Phi^{-1}(0.975)\hat{\sigma}_{\hat{\theta}},\hat{\theta}+\Phi^{-1}(0.975)\hat{\sigma}_{\hat{\theta}}]$, where $\Phi$ is
the probability function of the standard normal distribution and where
$\hat{\sigma}_{\hat{\theta}}$ are obtained from the inverse of the
estimated Hessian matrix, are also indicated.

\begin{figure}

\includegraphics{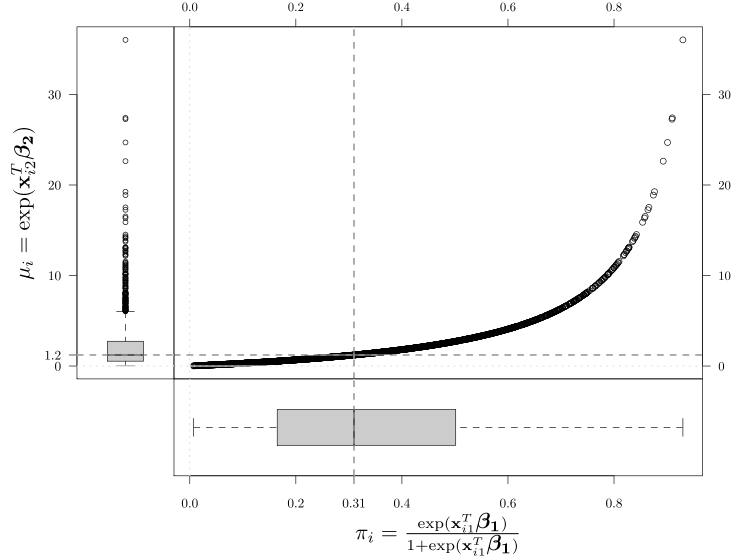}

\caption{Distribution of the probabilities of positive
outcomes, $\pi_{i}$, and of the expectations of positives values,
$\mu_{i}$, used to simulate ZITPo realizations.}\label{fig:simul2pimu}
\end{figure}

The data are simulated from a ZITPo distribution with parameters
\[
\pi_{i}=\frac{\exp(\mathbf{x}_{i1}^{T}\betab_{1})}{1+\exp(\mathbf{x}_{i1}^{T}\betab_{1})}\quad
\mbox{and}\quad \mu_{i}=\exp(\mathbf{x}_{i2}^{T}\betab_{2}).
\]
For the
covariates, the same $\mathbf{X}$ matrix is used in both parts of the
model. The first column of $\mathbf{X}$ is a column vector of $1$
corresponding to the constant. The other columns of $\mathbf{X}$ were
constructed with random values of respectively a normal, a Poisson, two
binomials and an exponential distribution, with corresponding
regression parameters $\betab_{1}=[1,1,-0.5,0.5,0.25,0.25]^{T}$ and
$\betab_{2}=[2,1,0.5,0.5,0.25,0.25]^{T}.$ The values of the
$\betab_{1}$ and $\betab_{2}$ were chosen in order to obtain
asymmetrical distributions for the probabilities of positive outcomes,
$\pi_{i}$, and for the expectations of positives values, $\mu_{i}$, as
well as a positive relationship between these quantities.
Figure~\ref{fig:simul2pimu} shows their respective distributions as
well as the chosen relationship between $\pi_{i}$ and $\mu_{i}$. With a
median of 0.3, the probabilities of positive outcomes, $\pi_{i}$, are
rather low. The expectations of the positives values, $\mu_{i}$, have a
very asymmetrical distribution. The cutting value $y^{\circ}=0.125$ is
a fixed value independent of $i$ and which approximately corresponds to
the quantile 0.1 of the positive simulated data. The form of the
dependance between $\pi_{i}$ and $\mu_{i}$ is nonlinear. The choice of
the parameter values $\pi_{i},\mu_{i}$ and $y^{\circ}$ thus corresponds
to an extreme choice to test the performance of the MLE in nontrivial
situations.

\begin{figure}

\includegraphics{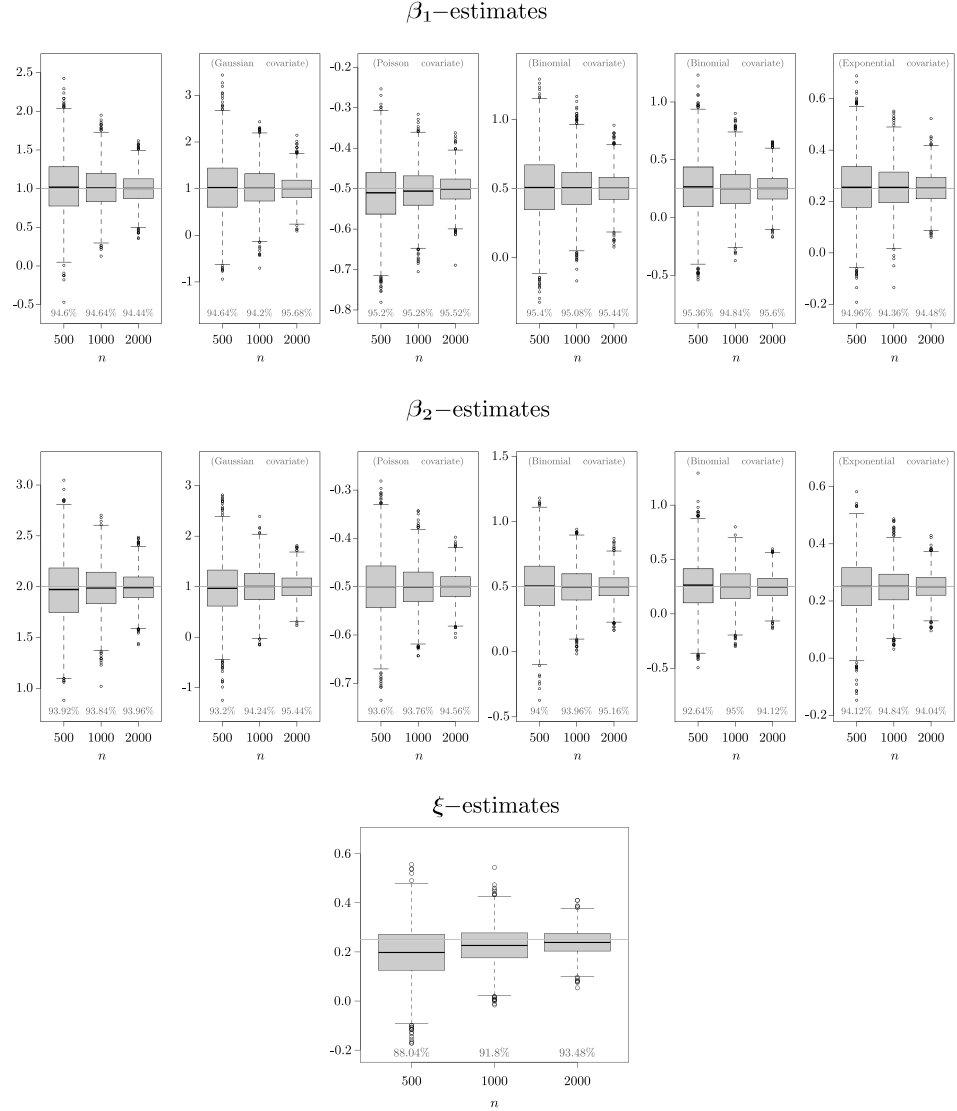}

\caption{Boxplots of the MLE of $\betab_{1}$ (upper
plots), $\betab_{2}$ (centered plots) and $\xi$ (bottom plot) computed
over 2500 datasets simulated from a ZITPo distribution with parameters
$\betab_{1}=[1,1,-0.5,0.5,0.25,0.25]^{T}$,
$\betab_{2}=[2,1,0.5,0.5,0.25,0.25]^{T}$ and $\xi=0.25$.
$y^{\circ}=0.125$ is a fixed value which approximately corresponds to
the quantile 0.1 of the positive simulated data. Analyses were
performed for samples of sizes 500, 1000 and 2000. The horizontal gray
lines indicate the position of the true parameter values. The coverage
levels of confidence intervals of the form
$[\hat{\theta}-\Phi^{-1}(0.975)\hat{\sigma}_{\hat{\theta}},\hat
{\theta}+\Phi^{-1}(0.975)\hat{\sigma}_{\hat{\theta}}]$,
where $\Phi$ is the probability function of the standard normal
distribution, are also indicated.} \label{fig:simul2beta}
\end{figure}

The bottom plot of Figure~\ref{fig:simul2beta} shows the sampling
distribution of the MLE of the shape parameter $\xi$. The boxplots of
the parameters estimates of $\xi$ show a small underestimation of the
parameter value even when the number of positive data is around 650
observations which correspond to 30\% of the maximum sample size of
this analysis. Estimation of the shape parameter is known to be
problematic even with large sample sizes and regardless of the
estimating method [\citet{HoskWall87}]. Our simulations tend to show that
the bias of the shape parameter both depends on the number of
observations $n$ and on the number of covariates $p$, a situation
similar to the MLE of the parameter $\sigma$ in multiple regression
analyses. This also confirms the findings of
Chavez-Demoulin and Davison [(\citeyear{DavisonChavez05}), page~212] for $\xi$ in their adaptation of
generalized additive models to the generalized Pareto distribution.

The upper and centered plots of Figure~\ref{fig:simul2beta} present
the sampling distributions of the MLE of $\betab_{1}$ and $\betab
_{2}$. Regardless of the sample size, all boxplots are well centered
around the true value of the parameters and the coverage levels of the
corresponding confidence intervals are close to the $95\%$ nominal
value. As $\betab_{2}$ and the $\xi$ are essentially estimated over
the positive part of the data which represent the 30\% of the 500, 1000
and 2000 observations of our study, our results appear very
satisfactory. Similar results were obtained in simulations with $\xi=0.5$.


\section{Model validation}\label{s:modelcheck}

Residual analyses in the context of models for data with excess zeros
as described in (\ref{eq:zerodiscr}) and (\ref{eq:zerocont}) may be
split in two parts: A first one focusing on the distribution that
distinguishes the zeros from the positive outcomes, and a second one
considering the distribution of the positive values. In models with
covariates, the residuals of the part distinguishing the zeros
correspond to residuals of logistic regressions. As this topic is
already well covered in the literature [we refer to \citet{Collet-03}
for a complete overview], the following subsections focus on the
residuals of the positive part of the model. We propose a residual type
for truncated and untruncated generalized Pareto models.
Section~\ref{s:application} presents one use of this new residual type.


Let $Y_{i}^{*+}=(Y_{i}^{*}|Y_{i}^{*}>0)$ denote the positive values of
the model and let $Y_{i}^{+}=(Y_{i}|Y_{i}>y^{\circ})$ be the observed
truncated positive values. As $(Y_{i}^{*+}-y^{\circ
}|Y_{i}^{*+}>y^{\circ}) = (Y_{i}^{+}-y^{\circ})$ and follows a
$\mathrm{GPD}(\mu_{i}+\frac{\xi y^{\circ}}{1-\xi},\xi)$, let us define
the $i$th residual, $\varepsilon_{i}$, in the following way:
%
\begin{equation}\label{eq:resid1}
\varepsilon_{i}=h(Y_{i}^{+}-y^{\circ})=\frac{Y_{i}^{+}-y^{\circ
}}{\mathrm{E}[Y_{i}^{+}-y^{\circ}]}=\frac{Y_{i}^{+}-y^{\circ}}{\mu_{i}
+ \xi y^{\circ}/(1-\xi)}.
\end{equation}
%
The residuals distribution, $f_{\varepsilon_{i} } (\varepsilon_{i})$, may
then easily be derived and is given by
\begin{eqnarray}\label{eq:resid2}
f_{\varepsilon_{i}}(\varepsilon_{i})
&=& f_{(Y^{+}_{i}-y^{\circ})}(h^{-1}(\varepsilon_{i})) \biggl| \frac{\partial
}{\partial\varepsilon_{i}} h^{-1}(\varepsilon_{i}) \biggr|\nonumber\\[-8pt]\\[-8pt]
&=& \frac{1}{1-\xi}\biggl(1+\frac{\xi}{1-\xi}
\varepsilon_{i}\biggr)^{-1/\xi-1}.\nonumber
\end{eqnarray}
Thus, $f_{\varepsilon_{i}}(\varepsilon_{i}) \sim\mathrm{GPD}(\mu=1,\xi)$. The
residuals theoretically (i.e., if the ZITPo model holds) follow a
generalized Pareto distribution of parameters $\mu=1$ and $\xi$. This
result holds also when $y^{\circ}=0$. Note that this result is a finite
sample result, a pretty rare situation in GLM.
A very powerful finite sample model validation procedure thus consists
in comparing the distribution of the
estimated residuals to their estimated theoretical distribution. The
former are obtained by substituting in (\ref{eq:resid1}) the parameters
by their estimated values, that is,
\[
\hat{\varepsilon}_{i}=\frac{Y_{i}^{+}-y^{\circ}}{\hat{\mu}_{i} +
\hat{\xi} y^{\circ}/(1-\hat{\xi})} \sim
\mathrm{GPD}(\mu=1,\hat{\xi}).
\]
QQ-plots should approximately display a straight line when the model
adequately fits the data.

Finally, the result in (\ref{eq:resid2}) offers a fast method to
generate random realizations from truncated or untruncated generalized
Pareto models. Indeed, let $u$ be a random realization of a
$\mathrm{Uniform(0,1)}$ and let $\mu$ be the vector of expectations of
the generalized Pareto distribution. Then, inverting (\ref{eq:resid1})
and (\ref{eq:resid2}) allows to generate $y$, a random variate of a
$y^{\circ}$-truncated $\mathrm{GPD}(\mu,\xi)$, in the following way:
\[
y=\biggl[(u^{-\xi}-1)\frac{1-\xi}{\xi} \biggr]
\biggl(\mu+\frac{\xi y^{\circ}}{1-\xi} \biggr)+y^{\circ}.
\]


\section{Applications to radio audience data}\label{s:application}

\begin{figure}

\includegraphics{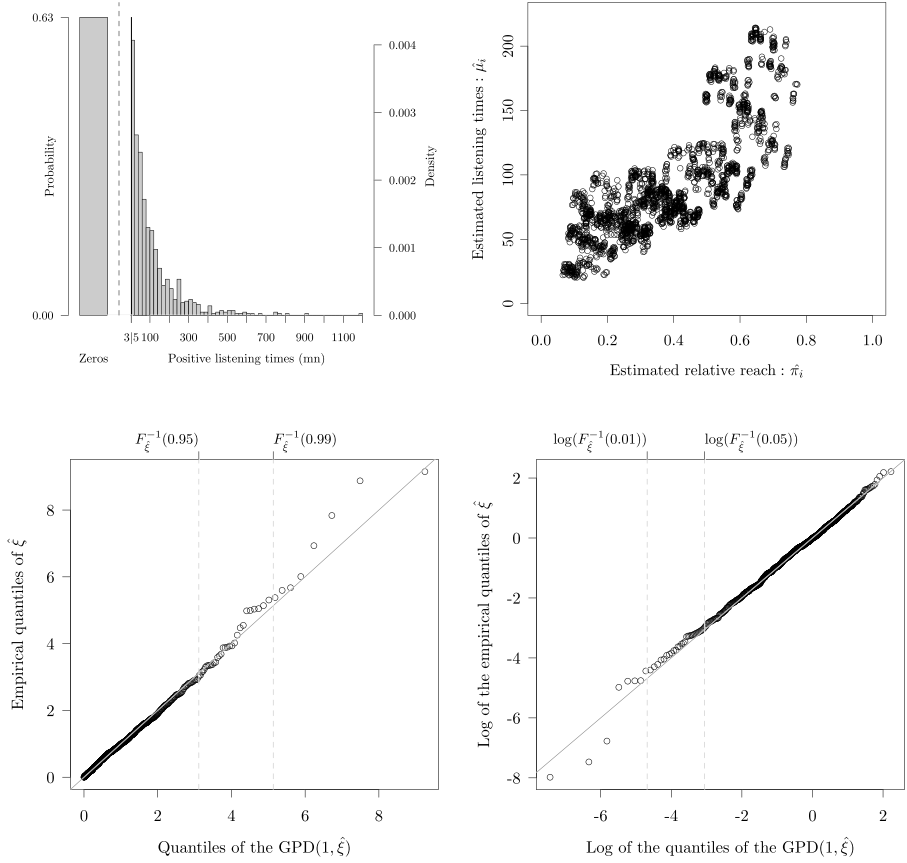}

\caption{Analyses of listening times to radio ``116'' in
its broadcasting area during the second semester of 2007. Left upper
plot: Empirical distribution function. The number of observations is
2155. The clump-at-zero represents the 63\% of the data. Right upper
plot: Form of the link between the estimated average ratings,
$\hat{\pi}_{i}$, and the average listening times, $\hat{\mu}_{i}$. Two
bottom plots: QQ-plots of the residuals \textup{(left)} and of the log of the
residuals \textup{(right)} of the ZITPo model applied to those data. The ordered
residuals are compared to the quantiles \textup{(left)} and to the log of the
quantiles \textup{(right)} of a $\mathrm{GPD}(1,\hat{\xi}= 0.082)$.}\label{fig:zitpo116B}
\end{figure}

The ZITPo model is applied to the audience data of the local radio
station ``116'' in its broadcasting area during the weekdays of the
second semester of 2007. The data set is available in
\citet{DLCMPVF10}. The left upper plot of Figure~\ref{fig:zitpo116B}
presents the distribution of the daily listening times of 2155
participants measured during one day of this period. The clump-at-zero
represents 63\% of the data.

The audience indicators of rating and time spent listening are
explained by a set of categorical variables including the age in 5
classes ($[15,25[,$ $[25,35[,$ $[35,45[,$ $[45,60[,$ $[60,120[$), the education
level in 3 classes (low, mid, high), the gender, the time in month and
the different zones of the broadcast area. The contrasts used to create
the $k-1$ dummy variables from a $k$-classes categorical variable are
of type ``treatment'' for the variables age, gender and education with
base ``15 to 25 years old men with low education level,'' and of type
``Sum'' for the geographical zones and the months. The model includes
interaction between age and gender. Other interactions---like between
education and age---appeared nonsignificant and did not improve the
log-likelihood or the residual distribution.

To protect the parameter estimates of the possible influence of the
false positive and false zeros observations belonging to the interval
$[0,5[$, we choose $y^{\circ}=4.95$. Consequently, we coded the 19
observations belonging to the interval $[3,5[$ in Figure~\ref
{fig:zitpo116B} as zeros and let the ZITPo model adequately separate
the true from the false zeros as described in the first part of~(\ref
{eq:semigpdmui}).


\begin{table}
\caption{$\betab_{1}$ and $\betab_{2}$ estimated
parameters and corresponding standard deviations of the ZITPo model
applied to the listening times to radio station ``116.'' The
$p$-values are for (asymptotic) significance testing of $\betab_{1}$
and $\betab_{2}$. Low $p$-values are magnified in the columns ``Sig.''
by means of (***), (**), (*), ($\cdot$) respectively corresponding to
significant tests at the levels of 0.001, 0.01, 0.05 and 0.1}\label{tab:zitpo116}
\begin{tabular*}{\textwidth}{@{\extracolsep{\fill}}lcccccccc@{}}
\hline
 &\multicolumn{4}{c}{\textbf{Rating}}&\multicolumn{4}{c@{}}{\textbf{Average listening time}} \\
 &\multicolumn{4}{c}{\hrulefill}&\multicolumn{4}{c@{}}{\hrulefill}\\
 &$\bolds{\hat{\beta}}_{\mathbf{1}}$&\textbf{SE}&\textbf{\textit{p}-value}&\textbf{Sig.}&
$\bolds{\hat{\beta}}_{\mathbf{1}}$&\textbf{SE}&\textbf{\textit{p}-value}&\textbf{Sig.}\\
\hline
(Intercept)&$-$1.95&0.32&$<$0.001&***&\phantom{$-$}4.08&0.33&$<$0.001&***\\
$[$25--35$[$&\phantom{$-$}0.40&0.39&\phantom{$<$}0.309&&$-$0.16&0.39&\phantom{$<$}0.680&\\
$[$35--45$[$&\phantom{$-$}0.94&0.36&\phantom{$<$}0.008&**&\phantom{$-$}0.20&0.35&\phantom{$<$}0.568&\\
$[$45--60$[$&\phantom{$-$}1.57&0.34&$<$0.001&***&\phantom{$-$}0.40&0.34&\phantom{$<$}0.235&\\
$[$60--120$[$&\phantom{$-$}2.22&0.35&$<$0.001&***&\phantom{$-$}0.76&0.34&\phantom{$<$}0.026&*\\
Women&$-$0.25&0.49&\phantom{$<$}0.608&&$-$0.73&0.49&\phantom{$<$}0.133&\\
Educ. middle&\phantom{$-$}0.18&0.16&\phantom{$<$}0.255&&\phantom{$-$}0.01&0.13&\phantom{$<$}0.933&\\
Educ. high&\phantom{$-$}0.36&0.12&\phantom{$<$}0.002&**&$-$0.15&0.09&\phantom{$<$}0.103&\\
July&$-$0.15&0.12&\phantom{$<$}0.216&&$-$0.06&0.10&\phantom{$<$}0.516&\\
August&$-$0.11&0.11&\phantom{$<$}0.346&&\phantom{$-$}0.04&0.09&\phantom{$<$}0.695&\\
September&\phantom{$-$}0.14&0.11&\phantom{$<$}0.225&&$-$0.07&0.09&\phantom{$<$}0.393&\\
October&\phantom{$-$}0.18&0.11&\phantom{$<$}0.085&$\cdot$&\phantom{$-$}0.01&0.08&\phantom{$<$}0.948&\\
November&$-$0.00&0.11&\phantom{$<$}0.973&&\phantom{$-$}0.04&0.09&\phantom{$<$}0.654&\\
Zone 2&$-$0.26&0.05&$<$0.001&***&$-$0.08&0.04&\phantom{$<$}0.049&*\\
Women $+ [$25--35$[$&$-$0.02&0.58&\phantom{$<$}0.970&&\phantom{$-$}1.26&0.57&\phantom{$<$}0.028&*\\
Women $+ [$35--45$[$&\phantom{$-$}0.18&0.54&\phantom{$<$}0.737&&\phantom{$-$}0.71&0.53&\phantom{$<$}0.180&\\
Women $+ [$45--60$[$&\phantom{$-$}0.03&0.52&\phantom{$<$}0.961&&\phantom{$-$}0.90&0.51&\phantom{$<$}0.079&$\cdot$\\
Women $+ [$60--120$[$&\phantom{$-$}0.39&0.52&\phantom{$<$}0.455&&\phantom{$-$}1.11&0.50&\phantom{$<$}0.027&*\\
\hline
\end{tabular*}
\end{table}

\begin{figure}

\includegraphics{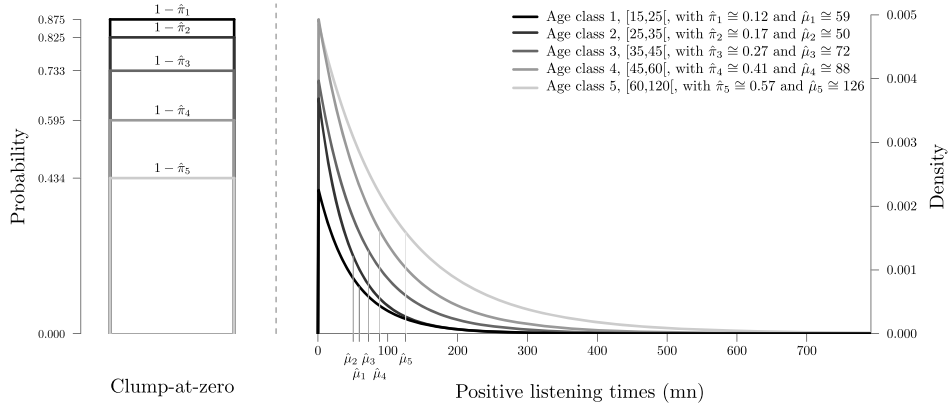}

\caption{Estimated (untruncated) listening times
distributions of men with low eduction level conditional to 5 age
classes. The probability to tune into this radio station strongly
depends on the age class. The expected listening times are more or less
the same except for the oldest age class.} \label{fig:zitpo116A}
\end{figure}

The $\betab_{1}$ and $\betab_{2}$ estimated values as well as their
standard deviations are reported in Table~\ref{tab:zitpo116}. The
$p$-values corresponding to the (asymptotic) significance tests for
$\betab_{1}$ and $\betab_{2}$, that is, $2\Phi^{-1}(-|\hat{\beta}/
\hat{\sigma}_{\hat{\beta}}|)$, are also indicated. According to the
chosen contrasts, the estimated intercepts $\beta_{10}$ and $\beta
_{20}$ are related to the estimated rating and time spent listening of
15 to 25-year-old men with a low education level in the broadcast area
of interest during the second semester of 2007 through, respectively,
\[
\frac{\exp(\hat{\beta}_{10})}{1+\exp(\hat{\beta }_{10})} \cong
0.12\quad
\mbox{and}\quad \exp(\hat{\beta}_{20}) \cong59.
\]
15--25-year-old men living in the broadcast area of interest and
having a low education level have thus a probability of contact to
radio station ``116'' of 12\% and an average contact length of about 59
minutes during the second semester of 2007.
The estimated distribution of the effective (untruncated) positive
times of the individuals of this focus group is thus
\[
Y_{i}^{*}|(Y_{i}^{*}>0) \sim\mathrm{GPD}(59,\hat{\xi}\cong0.082).
\]
Thus, under the model, $F^{-1}_{(Y^{*}|Y^{*}>0)}(3|59,\hat{\xi})\cong
0.05$ and $F^{-1}_{(Y^{*}|Y^{*}>0)}(y^{\circ}|59,\hat{\xi})\cong0.09$
respectively represent for this focus group the estimation of the part
of effective positive data that is coded as zero by the Swiss
measurement system and the estimation of the part of the effective
positive data that was supposed truncated and coded as zero for the
estimation.
The average ratings and time spent listening of other focus groups---like women with a high education level---are then shifts of 12\% and
59 minutes. Figure~\ref{fig:zitpo116A}, for example, presents the
estimated (untruncated) listening times distributions of men with low
eduction level conditional to 5 age classes. The probability to tune
into this radio station strongly depends on the age class. The expected
listening times are more or less the same for 15--45-year-old men and
increase then for the two oldest age classes.

In order to test the significance of each factor (e.g., age), we use
the likelihood ratio test to compare nested models. Let $\bolds{\beta
}=[\bolds{\beta}^{T}_{(1)},\bolds{\beta}^{T}_{(2)}]^{T}$ be the vector of
the regression parameters. The LRT statistic can be used to test
hypotheses of the form $H_{0}\dvtx\bolds{\beta}^{T}_{(2)}=0$ against
$H_{1}\dvtx\bolds{\beta}^{T}_{(2)}\neq0$ [with $\bolds{\beta}^{T}_{(1)}$
unspecified] and is given by
\[
\mathit{LRT}=2[l(\bolds{\hat{\beta}}|\mathbf{y},y^{\circ},\mathbf{X}_{1},\mathbf{X}_{2})-l(\bolds{\dot{\beta}}|\mathbf{y},y^{\circ},\mathbf{X}_{1},\mathbf{X}_{2})],
\]
where $\bolds{\hat{\beta}}$ and $\bolds{\dot{\beta}}$ respectively
denote the full and reduced regression parameters MLE. The LRT
statistic follows a $\chi^{2}_{p-\dot{p}}$ distribution under the
null hypothesis, where $p$ and $\dot{p}$ are the number of parameters
of the full and reduced model.

\begin{table}
\caption{LRT statistics (with corresponding degrees of
freedom) and $p$-values for the marginal LRT applied to the listening
times to radio station ``116.'' Each variable (or variable plus
interaction) of the left column is tested in the binomial (Rating) and
truncated GPD (Average listening time) part of the model. Low
$p$-values are magnified in the columns ``Sig.'' by means of (***),
(**), (*), ($\cdot$) respectively corresponding to significant tests at the
levels of 0.001, 0.01, 0.05 and 0.1} \label{tab:lrt116}
\begin{tabular*}{\textwidth}{@{\extracolsep{\fill}}lcccccccc@{}}
\hline
&\multicolumn{4}{c}{\textbf{Rating}}&\multicolumn{4}{c@{}}{\textbf{Average listening time}}\\
&\multicolumn{4}{c}{\hrulefill}&\multicolumn{4}{c@{}}{\hrulefill}\\
&\textbf{\textit{T}}&\textbf{Df}&\textbf{\textit{p}-value}&\textbf{Sig.}&\textbf{\textit{T}}&\textbf{Df}&\textbf{\textit{p}-value}&\textbf{Sig.}\\
\hline
Age $+$ age${}\cdot{}$gender&236.58&8&$<$0.001&***&78.54&8&$<$0.001&***\\
Gender $+$ age${}\cdot{}$gender&\phantom{00}3.26&5&\phantom{$<$}0.659&&16.08&5&\phantom{$<$}0.007&**\\
Education&\phantom{00}9.61&2&\phantom{$<$}0.008&**&\phantom{0}3.41&2&\phantom{$<$}0.182&\\
Month&\phantom{00}5.91&5&\phantom{$<$}0.315&&\phantom{0}1.53&5&\phantom{$<$}0.909&\\
Zone&\phantom{0}24.67&1&$<$0.001&***&\phantom{0}3.92&1&\phantom{$<$}0.048&*\\
Age${}\cdot{}$gender&\phantom{00}2.56&4&\phantom{$<$}0.634&&\phantom{0}8.14&4&\phantom{$<$}0.087&$\cdot$\\
\hline
\end{tabular*}
\end{table}

Table~\ref{tab:lrt116} presents the LRT evaluating which variables
significantly influence the rating and the average listening times.
According to the corresponding $p$-values, the variables significantly
influencing the average rating of this radio station are the age, the
education level and the geographical zone in the broadcast area. A~look
at the $\betab_{1}$ estimates shows that the rating average increases
with age and education classes and decreases for people living in the
countryside area named ``Zone~2.'' The variables significantly
influencing the average listening time to this radio station are the
age, the gender and area. The listening time average increases for
people belonging to high age classes and decreases for people living in
``Zone 2.'' The evolution of listening time with age is not the same
for men and women. The right upper plot of Figure~\ref{fig:zitpo116B}
shows the form of the link between the estimated average ratings, $\hat
{\pi}_{i}$, and the average listening times, $\hat{\mu}_{i}$: this
relationship is approximately linear, strong and positive (the correlation is of $0.78$).

The estimated shape parameter is $\hat{\xi}= 0.082$ with $\hat
{\sigma}_{\hat{\xi}}=0.039$. The shape parameter is thus slightly but
significantly higher than zero, meaning that a GLM with an exponential
error distribution, a special case of the ZITPo models when $\xi\to0$,
would not have been convenient in this case. The residuals are to be
compared to a $\mathrm{GPD}(1,0.082)$. The analysis of the fit is
presented in the two bottom plots of Figure~\ref{fig:zitpo116B}. The
QQ-plots of the residuals and of their log show a very good adequacy of
the model to the data.

Such conclusions represent a substantial improvement upon the available
ratings analyses in which point estimations of audience indicators are
calculated for the desired focus groups mostly without confidence intervals
and without the possibility to test the importance of a variable
compared to others. This information thus allows radio stations to
properly adapt their programming to better correspond to their desired
target audience, and advertisers to optimize targeted advertising campaigns.

\section{Conclusion}\label{s:conclusion}

The ZITPo model is a very powerful model that can be used, in
particular, to analyze radio audience data. Using the truncated
observations, this model allows to adequately estimate the true
proportions of nonzero observations and the average of positive values---corresponding to the audience indicators of rating and time spent
listening---of the underlying untruncated listening times
distribution. The model also allows to relate these expectations to
covariates in a GLM spirit, providing an explanatory model to audience
data. The model validation procedure resulting from properties of the
generalized Pareto distribution offers a very helpful way to judge the
adequacy of the model to the data.

Although the main motivation for the development of the ZITPo model was
the analysis of radio audience data, we believe that it can adequately
fit a number of data sets which have heavy tails distributions. For
example, it provides an extension to model (\ref{eq:singh}) for
hydrological data, that can include covariates to explain the mean
level, with $y^{\circ}=0$.


\section*{Acknowledgments}
The authors thank the Editor, an Associate Editor, a referee and E.
Cantoni for very constructive comments which greatly improved the
original manuscript. Radio data, measured by the Radiocontrol
measurement system, were kindly provided by the Mediapulse
Corporation.\footnote{\url{http://www.mediapulse.ch/en/home.html.}}


\begin{supplement}
\label{suppA} \stitle{Radio data set and R Code} \slink[doi]{10.1214/10-AOAS358SUPP}
\slink[url]{http://lib.stat.cmu.edu/aoas/358/supplement.zip} \sdatatype{.zip}
\sdescription{The file ``data\_ZITPo.csv'' contains the data set
analyzed in Section~\ref{s:application}. The observations are in rows
and the variables in columns. The file ``functions\_ZITPo.r'' contains
R functions that allow to fit and analyze ZITPo models. It produces
objects of class ``zipto.'' Usual generic functions are then available
for objects of that class. The file ``script\_ZITPo.r'' contains the R
Code used to produce the results of Tables \ref{tab:zitpo116} and~\ref{tab:lrt116} and the plots of Figure~\ref{fig:zitpo116B}.}
\end{supplement}


\printaddresses

\end{document}